\documentclass{wiley-article}

\usepackage{setspace}

\usepackage[round,authoryear,semicolon]{natbib}
\makeatletter
\renewcommand{\@biblabel}[1]{#1.}
\makeatother

\usepackage[utf8]{inputenc}
\usepackage{siunitx}
\usepackage{multirow}
\usepackage{booktabs}
\usepackage{siunitx}
\usepackage{etoolbox}
\usepackage{longtable}
\usepackage{float}
\usepackage{graphicx}
\usepackage{caption}
\usepackage{subcaption}
\usepackage{amsmath}
\usepackage{amsfonts}
\usepackage[
	pdftex,
	colorlinks=true,
]{hyperref}

\papertype{Applications}

\title{Whombat: An open-source audio annotation tool for machine learning assisted bioacoustics}

\author[1]{Santiago Martínez Balvanera}
\author[2]{Oisin Mac Aodha}
\author[3,4]{Matthew J. Weldy}
\author[1]{Holly Pringle}
\author[1,5]{Ella Browning}
\author[1]{Kate E. Jones}

\affil[1]{Centre for Biodiversity and Environment Research, Department of Genetics, Evolution and Environment, University College London, London, WC1E 6BT, United Kingdom}
\affil[2]{School of Informatics, University of Edinburgh, Edinburgh, EH8 9AB, United Kingdom}
\affil[3]{Department of Forest Ecosystems and Society, College of Forestry, Oregon State University, Corvallis, OR 97331-5704, USA}
\affil[4]{Pacific Northwest Research Station, USDA Forest Service, Corvallis, USA}
\affil[5]{Bat Conservation Trust, Studio 15 Cloisters House, Cloisters Business Centre, 8 Battersea Park Road, London, SW8 4BG, United Kingdom}

\corraddress{Santiago Martínez Balvanera}
\corremail{santiago.balvanera.20@ucl.ac.uk}

\fundinginfo{CONACYT, Grant/Award Number: 2020-000017-02EXTF-00334}

\runningauthor{Martínez Balvanera et al.}

\begin{document}

\renewcommand\refname{References}

\begin{frontmatter}
	\maketitle
	\label{sec:abstract}
\begin{abstract}
    \doublespacing
	{\bf Abstract}

	\small
    \noindent{}1.
    Automated analysis of bioacoustic recordings using machine learning (ML) methods has the potential to greatly scale biodiversity monitoring efforts.
    The use of ML for high-stakes applications, such as conservation and scientific research, demands a data-centric approach with a focus on selecting and utilizing carefully annotated and curated evaluation and training data that is relevant and representative.
    Creating annotated datasets of sound recordings presents a number of challenges, such as managing large collections of recordings with associated metadata, developing flexible annotation tools that can accommodate the diverse range of vocalization profiles of different organisms, and addressing the scarcity of expert annotators.

    \noindent{}2.
    We present \texttt{Whombat} a user-friendly, browser-based interface for managing audio recordings and annotation projects, with several visualization, exploration, and annotation tools.
    It enables users to quickly annotate, review, and share annotations, as well as visualize and evaluate a set of machine learning predictions on a dataset.
    The tool facilitates an iterative workflow where user annotations and machine learning predictions feedback to enhance model performance and annotation quality.

    \noindent{}3.
    We demonstrate the flexibility of \texttt{Whombat} by showcasing two distinct use cases:
    (1) an project aimed at enhancing automated UK bat call identification at the Bat Conservation Trust (BCT), and
    (2) a collaborative effort among the USDA Forest Service and Oregon State University researchers exploring bioacoustic applications and extending automated avian classification models in the Pacific Northwest, USA.

    \noindent{}4.
    \texttt{Whombat} is a flexible tool that can effectively address the challenges of annotation for bioacoustic research.
    It can be used for individual and collaborative work, hosted on a shared server or accessed remotely, or run on a personal computer without the need for coding skills.
    The code is open-source, and we provide a user guide.

\end{abstract}

\end{frontmatter}

\section{Introduction} \label{sec:intro}

Recent advancements in Machine Learning (ML) are revolutionizing our ability to analyse large datasets generated by passive acoustic recorders for ecologically relevant signals \citep{kitzesExpandingNEONBiodiversity2021,tuiaPerspectivesMachineLearning2022a}.
Open-source Deep Learning models, such as BirdNET \citep{kahlBirdNETDeepLearning2021a} and NABat ML \citep{khalighifarNABatMLUtilizing2022a}, can be used to monitor birds and bats at scale across large regions.
While considerable attention has been directed towards developing sophisticated ML systems, it is crucial to acknowledge the pivotal role of data and data work in establishing reliable ML implementations \citep{nithyasambasivanEveryoneWantsModel2021}.
In line with this, the data-centric approach has gained increasing relevance \citep{jarrahimohammadhosseinPrinciplesDataCentricAI2022}, emphasizing the collection, curation, and management of high-quality training and evaluation data to comprehensively assess model performance and ensure reliability, particularly in high-stakes applications such as conservation.
Data work is inherently complex, and audio annotation, encompassing the identification of relevant sound events' location in audio recordings and the assignment of appropriate labels, represents a time-consuming and labor-intensive process \citep{cartwrightCrowdsourcingMultilabelAudio2019b}.
Often, the creation of ML-ready datasets relies on software tools and technical infrastructure to ease management and enhance efficiency \citep{reichertNABatTopdownBottomup2021,roeAustralianAcousticObservatory2021a}.
However, while the broader ML community has recognized the importance of providing accessible, efficient and open-source tools for dataset curation and annotation \citep{sager2021survey,nevesAnnotationsaurusSearchableDirectory2020}, the bioacoustics community has lagged behind \citep{stowell2022computational,tuiaPerspectivesMachineLearning2022a}.

The annotation process is an integral part of an iterative workflow aimed at continually improving and monitoring the performance of ML models and data quality \citep{hohmanUnderstandingVisualizingData2020}.
The evaluation of ML models can help identify errors and areas for potential improvement, such as annotation or data gaps, thereby increasing confidence in the model's performance \citep{nahar2022collaboration}.
Continual annotation of novel data is crucial to monitor the performance of ML models, particularly when exposed to unknown environments, as these can pose a risk to model accuracy and reliability \citep{sariaTutorialSafeReliable2019}.
However, existing annotation tools often lack appropriate design for effective annotation and machine learning development, hindering the seamless execution of this valuable feedback loop (Table~\ref{table:software_comparison}).

Creating ML-ready datasets for bioacoustic research is a collaborative effort \citep{zhangHowDataScience2020} that requires a combination of modelling, analysis, annotation work, and quality assurance \citep{jarrahimohammadhosseinPrinciplesDataCentricAI2022,mullerHowDataScience2019}.
Annotation can be accelerated if tackled by teams working simultaneously and distributing the workload among members with specialized and expert knowledge \citep{mullerDesigningGroundTruth2021,cartwrightCrowdsourcingMultilabelAudio2019b}.
However, managing large collections of audio recordings in bioacoustic research can be overwhelming \citep{kvsnBioacousticsDataAnalysis2020a} as they often contain hundreds or thousands of recordings \citep{zhangManagingAnalysingBig2013}, each with its own set of metadata such as location, date, and time of recording, as well as other relevant contextual information.
Storing the associated metadata is desired as it can influence modelling decisions and provide contextual cues for acoustic identification \citep{kshirsagarBecomingGoodAI2021,paullada2021data}.
Being able to locate specific recordings or annotations within these collections is crucial for effective analysis and research but can be time-consuming and difficult without proper tools \citep{kandelEnterpriseDataAnalysis2012}.
Providing a platform for collaborative annotation requires finding a balance between accessibility, simplicity, and the ability to manage complex and diverse workflows \citep{simpsonZooniverseObservingWorld2014}.

\begin{table*}[ht]
	\centering
	\begin{tabular}{lccccccc}
 & \textbf{Whombat} & Arbimon\textsuperscript{\citenum{aideRealtimeBioacousticsMonitoring2013}} & AvianZ\textsuperscript{\citenum{marslandAviaNZFutureproofedProgram2019a}} & Kaleidoscope\textsuperscript{\citenum{acousticsKaleidoscopeProAnalysis2019}} & \shortstack{Label \\ Studio\textsuperscript{\citenum{Label}}} & Raven\textsuperscript{\citenum{yangCornellLabOrnithology2023}} & \shortstack{Sonic \\ Visualiser\textsuperscript{\citenum{SonicVisualiser}}} \\
		\toprule
Open-source& \checkmark{} & – & \checkmark{}& –& \checkmark{}& – & \checkmark{} \\
		Self-Host& \checkmark{} & – & \checkmark{}& \checkmark{} & \checkmark{}& \checkmark{}& \checkmark{}\\
		Collaborative& \checkmark{} & \checkmark{}& – & –& \checkmark{}& – & – \\
		Large Datasets & \checkmark{} & \checkmark{}& \checkmark{}& –& \checkmark{}& – & – \\
		Rich Metadata& \checkmark{} & \checkmark{}& – & \checkmark{} & – & – & – \\
		Search Capabilities& \checkmark{} & \checkmark{}& – & –& – & – & – \\
        Annotation Exploration & \checkmark{} & \checkmark{} & – & –& – & – & – \\
		Flexible Spectrogram & \checkmark{} & – & \checkmark{}& \checkmark{} & – & \checkmark{}& \checkmark{}\\
        Flexible Annotations & \checkmark{} & \checkmark{} & \checkmark{}& –& – & – & – \\
        Quality Assurance& \checkmark{} & \checkmark{} & – & –& – & – & – \\
		Training Tools & \checkmark{} & – & – & –& – & – & – \\
		Prediction Evaluation& \checkmark{} & \checkmark{}& \checkmark{}& –& – & – & – \\
        Export Annotations& \checkmark{} & – & – & – & \checkmark{} & \checkmark{} & \checkmark{} \\
		Integrated Detectors & –& \checkmark{}& \checkmark{}& \checkmark{} & – & – & – \\
		\bottomrule
	\end{tabular}
\caption{
Comparison of seven popular software used for acoustic annotation (see Supporting Information~\ref{appendix:software_comparison} for further details).
Dashes indicate that the corresponding feature is not supported by the software (to the best of the authors' knowledge), while a checkmark indicates its availability.
}
	\label{table:software_comparison}
\end{table*}

Bioacoustic annotation is a challenging task due to the wide variety of organisms and vocalization profiles that are studied in bioacoustic research \citep{stowell2022computational,odomComparativeBioacousticsRoadmap2021}.
Some animals produce long duration and broad-band sounds, while others produce vocalizations that can be clearly localized both in time and frequency.
Substantial expertise in the target animal's sound indentification is often required and aquiring this knowledge can be a challenging process, often requiring extensive field experience.
The pool of bioacoustic experts per taxa is, therefore, typically small and their expert annotation time is valuable \citep{nahar2022collaboration}.
In order to effectively accomodate the varying characteristics of different types of bioacoustic sounds, annotation tools must be flexible in terms of their visualization and annotation capabilities \citep{stowell2022computational}.
Annotated datasets have the potential to increase the pool of annotators by providing multiple examples of sounds for training, but this capability has not been fully developed within existing annotation tools.
Furthermore, generic audio annotation tools are primarily focused on the analysis of human speech or music and and lack the necessary visual representation of audio and consideration of recording context.
Conversely, speciallized bioacoustic software has often focused on specific taxonomic groups \citep{szewczakSonoBat2010a,marslandAviaNZFutureproofedProgram2019a}, making it difficult to use these tools for the analysis of other groups.
Despite the avaliability of a variety of annotation tools, none have been able to fully address the complexity of challenges that are inherent to bioacoustic research (Table~\ref{table:software_comparison}; see Supporting Information~\ref{appendix:software_comparison} for a thorough evaluation of audio annotation tools).

\begin{figure*}[ht]
	\centering
	\includegraphics[width=0.9\textwidth]{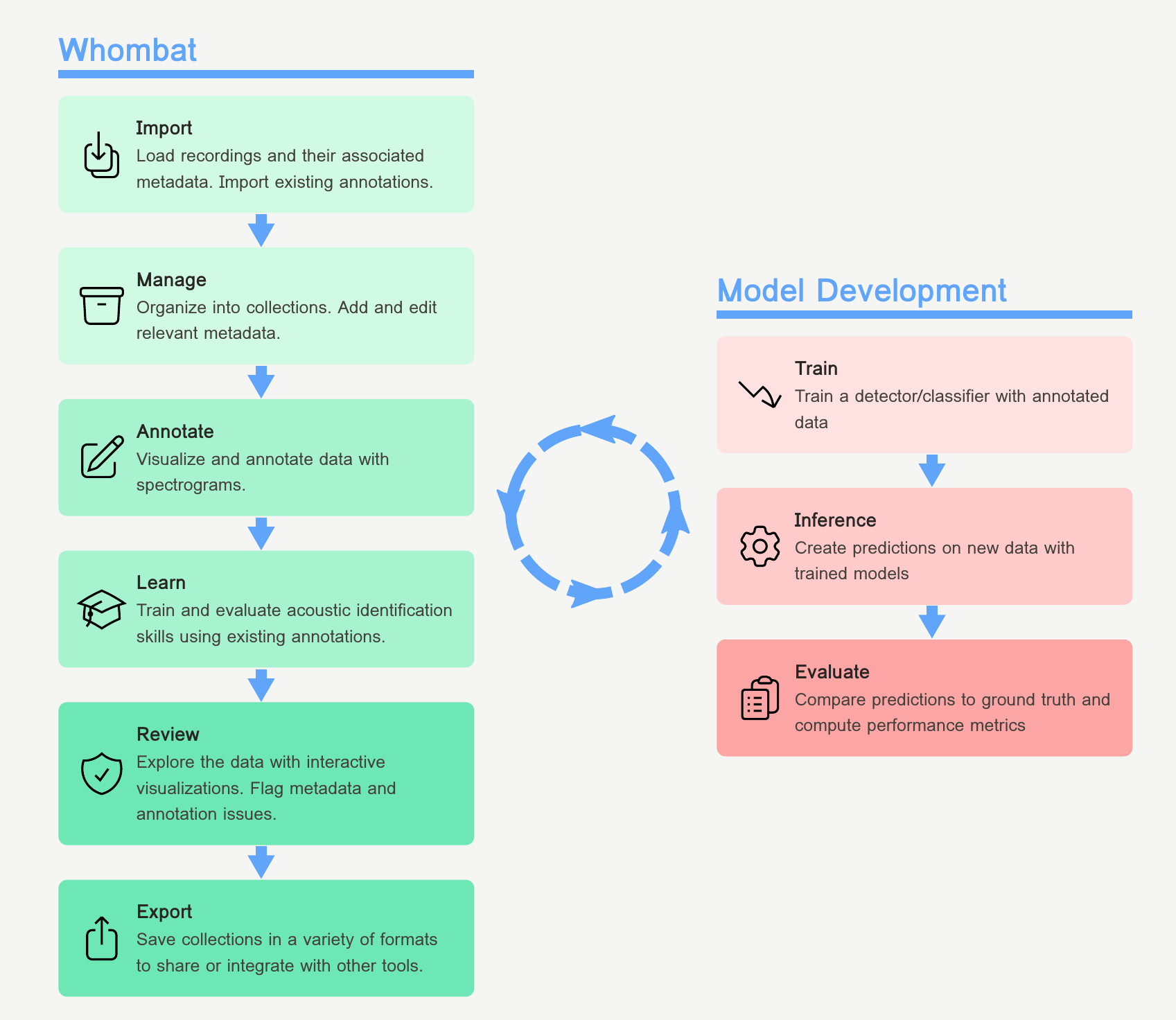}
	\caption{The iterative workflow of the \texttt{Whombat} annotation tool.
        The app enables a feedback loop between user annotations and machine learning predictions, enhancing both model performance and annotation quality.
        The capabilities of the tool are represented by the green boxes on the left, while the red boxes on the right illustrate the steps in the model development workflow.
        The arrows indicate the typical direction of the workflow, but the tool provides flexibility for users to navigate between steps.
    Dashed arrows indicate potential crossover between model and data development, providing users with the flexibility to explore and integrate both components as needed.}
	\label{fig:whombat_workflow}
\end{figure*}

Here we present \texttt{Whombat}, a flexible tool specifically designed to accelerate bioacoustic ML research by facilitating the curation of annotated acoustic datasets.
\texttt{Whombat} offers a user-friendly browser-based interface that enables efficient management of audio recording datasets and annotation projects.
It provides various visualization, exploration, and annotation tools that allow users to annotate, review, and share annotations with ease.
Moreover, these exploration tools can be employed to visualize, evaluate, and explore ML predictions on annotated datasets.
\texttt{Whombat} supports an iterative workflow (Fig.~\ref{fig:whombat_workflow}), where user annotations and ML predictions continuously enhance both model performance and annotation quality.
Additionally, \texttt{Whombat} is designed to support both individual and collaborative work, enabling hosting on shared servers, cloud platforms, or private premises with remote accessibility.
Notably, it can also run on personal computers without internet access.
The application code is open-source and available at \url{https://github.com/mbsantiago/whombat}.
To ensure accessibility for all users, we have bundled the tool into executable files for Windows, macOS, and Ubuntu, eliminating the need for dependency installation or coding skills.
By making \texttt{Whombat} open-source and easily accessible, we aim to empower researchers in bioacoustic ML research and foster advancements in the field.

\section{Whombat Features} \label{sec:application-features}

In this section we provide a brief description of \texttt{Whombat}'s features and interface, following the order of the intended annotation workflow (Fig~\ref{fig:whombat_workflow}).
This includes the initial setup and loading of data, visualization and navigation tools, annotation capabilities, quality control features, and ML model evaluation.
Through this overview, we demonstrate how \texttt{Whombat} can enhance the efficiency and accuracy of bioacoustic annotation.

\subsection{Dataset management} \label{sub:dataset-management}

The workflow begins by creating a dataset of acoustic recordings (Fig~\ref{fig:whombat_workflow}).
A dataset can be created by selecting all recordings within a folder or by importing a pre-existing dataset.
Currently, the tool only accepts audio files in WAV format.
Multiple datasets can be managed simultaneously.

\begin{figure*}[ht]
	\centering
	\begin{center}
		\includegraphics[width=\textwidth, height=10cm]{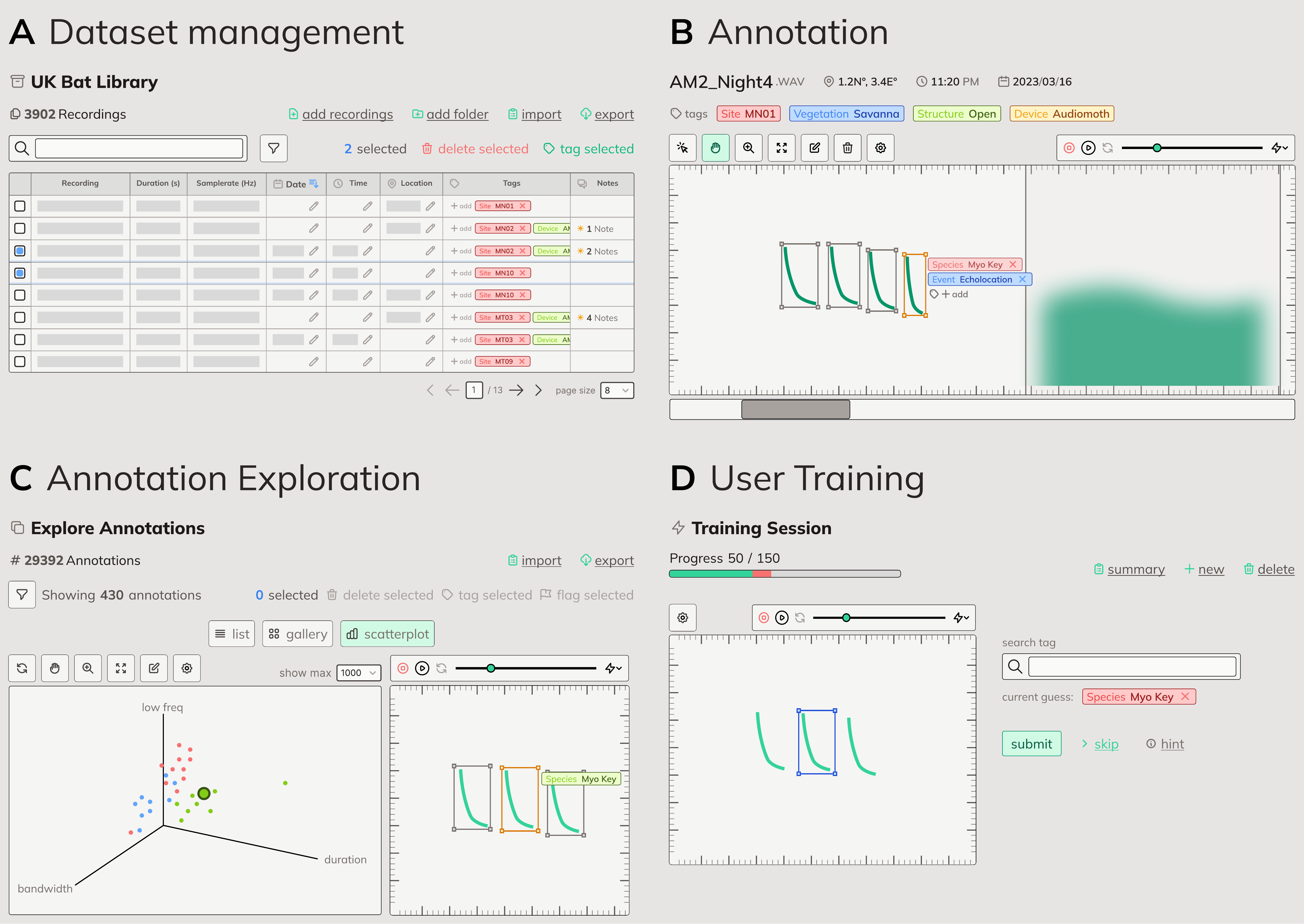}
	\end{center}
	\caption{
        Overview of the key features of \texttt{Whombat}.
        {\bf A} presents dataset management capabilities, including editing, searching, and filtering recording metadata.
        {\bf B} showcases the main annotation interface, offering spectrograms of recordings and various annotation tools.
        {\bf C} demonstrates annotation exploration features, enabling users to browse existing annotations through filtering and visualizations, such as scatter plots.
        Finally, {\bf D} highlights the training component of \texttt{Whombat}, where users learn to identify a collection of sound events while receiving guidance on sound event identification.
    }
	\label{fig:ui_panel}
\end{figure*}

Basic media information is scanned and stored for each recording, including its duration, number of channels, and sample rate (Fig.~\ref{fig:ui_panel}).
\texttt{Whombat} also allows the retrieval of metadata from some common acoustic hardware types (for example, Wildife Acoustics and AudioMoth \citep{hillAudioMothLowcostAcoustic2019a}).
Users can edit the location and date-time of recordings on a per-recording basis or import this information from CSV files.
Additionally, recordings can be tagged with multiple key-value pairs, providing contextual information relevant to the annotation process.

To explore datasets, users listen to recordings and visualize their spectrograms.
\texttt{Whombat} uses spectrograms as the main visualization tool as they facilitate quick identification of sound events \citep{cartwrightCrowdsourcingMultilabelAudio2019b}.
Spectrogram parameters and other visual settings are configurable to best suit target sounds.
\texttt{Whombat} dynamically generates spectrogram sections on the fly, optimizing computational efficiency and preventing excessive memory usage for long recordings. 
This allows for easy navigation using scroll bars, eliminating the need to compute and store large spectrograms in their entirety.
Users can zoom in to relevant parts of the spectrogram or zoom out to scan for interesting sounds.
\texttt{Whombat} also provides searching, filtering, and sorting tools to quickly browse the recordings of interest.

\subsection{Annotation} \label{sub:annotation}

Annotation projects can be created by selecting any number of audio clips from recordings of interest.
Here audio clips are contiguous fragments of recordings of any length.
The use of audio clips as the basis of annotation tasks allows cutting the recordings into clips of standardised duration and possibly annotating only a subset of all audio clips.
The included clips can be selected within the tool or imported from a CSV file.
To create an annotation project, a name, description, and annotation instructions for the annotators should be provided.

Once an annotation project is created, each audio clip can then be visualised and annotated.
A configurable spectrogram of the clip is displayed, along with recording metadata to provide context to the annotator (Fig.~\ref{fig:ui_panel}).
Annotation can proceed in different ways depending on the project targets and strategy.
Users can add any number of key-value tags to the recording clip, for example to specify which species are present within the clip.
Relevant sound events can be annotated by locating them within the spectrogram by drawing a vertical line, a temporal interval or a bounding box.
Each annotation can be tagged with any amount of key-value pairs, potentially capturing multiple and independent attributes of the sound event, such as species, sound type, sex, or the identity of the individual.
Although tags can be created freely, we offer a quick search feature to avoid duplication and to ensure consistency.

As annotations progresses, audio clips can be marked as ``ready'' once they have been fully annotated according to the project instructions.
Annotation progress is tracked by displaying the percentage of audio clips that have been marked as ready, along with the counts of annotated clips and annotations with a given tag.
With the aid of filtering and sorting tools, users can focus and prioritise their annotation efforts on specific subsets of the annotation project.

\subsection{Review and exploration} \label{sub:review-and-exploration}

Quality of metadata and annotations can be reviewed and managed through various tools within \texttt{Whombat}.
Users can add notes to recordings and annotations to provide additional context and note issues that require fixing.
Incomplete annotations can also be flagged, and the issues can be searched to address them efficiently.

In addition, \texttt{Whombat} provides tools for exploring and comparing groups of annotations.
The gallery option displays a panel of annotated sound events from different user-selected groups, allowing for easy comparison.
For groups of bounding box annotations, the tool can compute statistics on attributes such as the duration, bandwidth, and frequency range, and display them in histograms.
\texttt{Whombat} also provides an interactive 2D or 3D scatter plot of any combination of the previously mentioned attributes (Fig.~\ref{fig:ui_panel}).
These visualisation tools enable users to become familiar with the variety of sound events and identify potential issues such as outliers and overlaps between categories.

\subsection{User training} \label{sub:user-training}

Novice users can be incorporated into the workflow by training with existing verified data, which is critical as access to experts in bioacoustics is a recurrent bottleneck for annotation.
Our tool addresses this issue by allowing users to learn and improve their annotation skills. The registered annotations can be used to train and evaluate human annotation skills.
Users can create training sets by selecting specific annotations, such as those with a particular set of tags.
The training sets can be used to conduct training sessions (Fig.~\ref{fig:ui_panel}), in which users are shown a series of spectrograms centred at annotated sound events and asked to identify them correctly.
After each session, the identification performance is evaluated and displayed, enabling users to track their learning progress and identify areas that need improvement.

\subsection{Data export} \label{sub:data-export}

\texttt{Whombat} allows users to export their recording datasets and annotation projects to multiple file formats.
The recommended format is a custom JSON format inspired by the COCO dataset format \citep{lin2014microsoft}, although a CSV format is also available.
This makes it possible to use the annotations for training machine learning identification models, other bioacoustic analysis, or to share with the wider community.

In addition, the exported dataset and annotation project files can be imported back into the tool.
This functionality allows for offline distributed collaborative work where multiple people work on disjointed datasets and share the resulting annotations.
This is particularly useful because it bypasses the need for centralized server infrastructure.

\subsection{Closing the loop} \label{sub:closing-the-loop}

To improve ML model performance, the tool provides a way to import model predictions and compare them with user-made annotations.
\texttt{Whombat} accepts model predictions in a specific JSON or CSV format, with no restriction on the type of ML model used.
Once imported, the set of predictions for a group of recordings is registered as a model run.
Users can provide a name and description for the run to help track and organize different model experiments.

\texttt{Whombat} then allows users to evaluate the model run by comparing it with annotations, if available.
Several summary detection metrics are displayed, such as precision, recall, to help users assess the model's performance.
Users can also explore the predictions using search, sort, and filter tools, based on the predicted tag probabilities.
This facilitates browsing both success and failure cases, helping to identify potential model improvement opportunities.

In addition to evaluating the ML model, the tool can also be used to diagnose potential data and annotation gaps.
By comparing the model predictions with user-made annotations, users can identify cases where the model fails to detect sound events correctly.
These cases can then be reviewed to see if there are annotation or data gaps that need to be addressed to improve model performance.

\section{Use Cases} \label{sec:use-cases}

\texttt{Whombat} is designed specifically for audio data in the field of bioacoustics, and its flexibility makes it adaptable to a range of use cases.
In this section, we highlight two examples of how the tool can be used: bat call annotation of UK bats species and bird vocalization detection in the Pacific Northwest of the USA\@.
These examples showcase the versatility and potential of the tool for annotating different types of target species and vocalisations.

\subsection{Bat call classification pipeline}

The Bat Conservation Trust (BCT) uses \texttt{Whombat} to improve bat detection and classification in the UK.
The BCT collects and annotates recordings of bat calls across the UK to enhance the BatDetect tool \citep{macaodhaBatDetectiveDeep2018} and advance from bat call detection to a multi-class object detection and classification pipeline \citep{macaodhaGeneralApproachBat2022}.
Bats play a crucial role in the UK ecosystems \citep{barlowCitizenScienceReveals2015}, and as small, nocturnal, volant mammals that use of ultrasonic echolocation for navigation are routinely monitored using passive acoustic methods \citep{bannerImprovingGeographicallyExtensive2018a, barlowCitizenScienceReveals2015, kerbiriouVigieChiroAnsSuivi2015, newsonNovelCitizenScience2015a, yohBenignEffectsLogging2023}.
Furthermore, interspecific differences in bat echolocation call characteristics enables species or genus level identification from acoustic data.
Automating the classification of bat echolocation calls enables monitoring to be carried out at the scales neccesary for identifying national conservation management strategies.
Improving the detection and classification performance of automated tools, such as BatDetect \citep{macaodhaGeneralApproachBat2022}, is therefore crucial for the success of conservation efforts.

\texttt{Whombat} facilitates simultaneous and collaborative annotation of bat calls by multiple users.
Bat calls are short and high-frequency, making them well-suited for annotation with bounding boxes tightly placed around the main harmonic.
Annotators use tags in the form \texttt{species}: \emph{<species>} to indicate the bat species, and \texttt{event}: \emph{<call type>} to specify the call type (e.g., echolocation, social call, feeding buzz).
In cases of uncertainty, a generic tag like \texttt{order}: \emph{Chiroptera} can be employed.
Additionally, potential false positives can be annotated with an \texttt{event}: \emph{Noise} tag to reduce confusion.

The tool has enabled the annotation of 29 independent datasets of bat recordings, yielding over 70,000 annotated calls.
It has been used by more than 15 independent annotators at the BCT and partner institutions.
The annotations are helping refine the detection and classification capabilities of the ML algorithms, with the aim to enhance understanding of bat population responses to anthropogenic environmental change and inform conservation efforts.
The collaborative nature of the tool also allows for efficient data sharing and analysis, making it an essential tool for BCT and their bat conservation work.

\subsection{Bird song annotation}

In 1994, the Northwest Forest Plan was introduced in the United States Pacific Northwest to shift federal land management policies from prioritizing timber harvesting to a more holistic approach that includes protecting and restoring the habitat of old-forest species and biodiversity \citep{espy1994record}.
One of the components of this plan is the long-term monitoring of federally threatened northern spotted owl (\textit{Strix occidentalis caurina}) populations through a two-phase approach \citep{lint1999northern}.
The first phase involved estimating vital rates and demographic performance using mark-resight methods on historical territories \citep{franklinRangewideDeclinesNorthern2021}.
The second phase began in 2020 and focused on estimating occupancy and habitat models through passive acoustic monitoring \citep{lesmeister2022integrating}. 

The transition to phase two monitoring is a crucial moment in conserving and managing forested lands in the Pacific Northwest.
Not only are spotted owl conservation and management objectives being met \citep{lesmeister2022integrating,weldy2023long}, but the multispecies acoustic monitoring data can also be used to address other conservation, research, or management objectives.
To this end, researchers from Oregon State University and the USDA Forest Service are using \texttt{Whombat} to annotate avian sounds (> 30,000 annotations) and validate model predictions for various wildlife monitoring programs targeting federally threatened species like the northern spotted owl and the marbled murrelet (\textit{Brachyramphus marmoratus}), as well as sensitive species such as the white-headed woodpecker (\textit{Dryobates albolarvatus}), and supporting broader biodiversity monitoring efforts (> 80 species).

The tools's dynamic acoustic and spectrogram adjustments have improved the quality of target species annotation, increased efficiency in reviewing model predictions, and aided in tracking acoustic review and labeling efforts.
In addition, the annotation formatting of the \texttt{Whombat} is flexible and dynamic, allowing annotators to pursue multiple annotation objectives simultaneously.
They can opportunistically collect biophonic examples for non-target species and create hierarchical label structures where sound types are nested within broader categories. 
These hierarchical labels cascade across increasingly fine-scale taxonomic determinations.
Additionally, annotators can label acoustic metadata such as approximate distances or overlapping sound types, which serves to improve model training and enhance the understanding of model performance.
The adoption of passive acoustic monitoring represents an important step forward in conserving and managing forested lands in the Pacific Northwest.
Using innovative tools such as \texttt{Whombat} enhances these efforts.

\section{Discussion} \label{sec:discussion}

The modularity and extensibility of \texttt{Whombat} enables many opportunities for future development (see the Supporting Information~\ref{appendix:software_design} for more details on the software design).
We invite the community to contribute to its growth and suggest potential areas of improvement, such as the ability to group annotations into sequences, model comparison and data iteration visualizations, and dashboards for ecological insights and quick exploration.
One possibility for expanding the tools's user base is to incorporate a citizen science approach by evaluating the reliability of user annotations.
We believe these and other potential directions will help make \texttt{Whombat} an even more powerful tool for bioacoustics research and conservation efforts.

Unlike other annotation solutions (e.g.~\cite{marslandAviaNZFutureproofedProgram2019a}), our tool does not include embedded machine learning detectors.
We made this decision to simplify the software and decouple the annotation process from the development and maintenance of machine learning models.
Instead, our focus is on providing a user-friendly interface for efficient and accurate annotation.
We also provide an interface for importing and exporting model predictions, allowing users to incorporate their own machine learning models into their annotation projects.
Additionally, the tool is able to export annotations in a format that is compatible with training frameworks for bioacoustic detection models (e.g.~\cite{macaodhaGeneralApproachBat2022}).

By providing an accessible, open-source tool for bioacoustic annotation, we hope to empower research teams to generate high-quality acoustic datasets for their projects, including those without extensive coding experience.
The modular and extensible design of the software allows for customization to meet individual project needs and encourages community involvement in the development of new features.
By lowering the barrier to entry for annotation projects, we aim to foster the creation of diverse and shareable datasets that can advance research in bioacoustics.

\section*{Acknowledgements}

We thank Giada Giacometti and other team members from the BCT, Sospeter Kiwibot and Paul Webala for beta testing and valuable feedback.
We also thank Grant Van Horn, Everardo Gustavo Robredo Ezquivelzeta, Veronica Guiterrez-Zamora and Cristina MacSwiney for helpful discussions.
Financial support was provided by Consejo Nacional de Ciencia y Tecnologia (CONACYT) Award Number 2020-000017-02EXTF-00334.

\section*{Conflict of Interest statement}

The authors declare that they have no conflict of interest.

\section*{Author Contribution}

OMA, SMB and KJ conceived the project and designed the software.
SMB wrote the software, the documentation and the frist draft of the manuscript.
EB, HP and MW led annotation efforts described in the use cases and provided extensive feedback on the software usability.
KJ and OMA reviewed the manuscript and provided feedback on the software design.
All authors contributed critically to the drafts and gave final approval for publication.

\section*{Code and Data Availability}

Whombat can be downloaded for Windows, macOS and Linux from~\url{https://github.com/mbsantiago/whombat/releases}.
The source code is available at~\url{https://github.com/mbsantiago/whombat}, and the README provides installation instructions.
The Software guide, which provides detailed usage instructions, is available in English at~\url{https://mbsantiago.github.io/whombat/}. Sample datasets mentioned in the article are included in the GitHub repository.

\bibliography{library}

\newpage
\section*{Supporting Information}

\appendix

\section{Software Comparison}
\label{appendix:software_comparison}

To obtain a comprehensive list of potential alternative tools to compare with \texttt{Whombat}, we conducted a thorough search using multiple sources.
Our search strategy included three main categories of sources.
First, we conducted searches in academic databases, specifically the Web of Science Core Collection and the IEEE Xplore Digital Library, for publications related to bioacoustic and audio annotation tools.
Second, we used search engines such as Google and GitHub to broaden our search.
Specifically, we conducted a Google search for ``audio annotation tool'' and ``bioacoustic software'' and searched on GitHub for public repositories with the tags ``audio'' and ``annotation.''
Finally, we consulted compiled lists of bioacoustic and annotation software.
Specifically, we referred to a GitHub repository maintained by \href{https://github.com/rhine3/bioacoustics-software}{rhine3}, which provides an up-to-date list of bioacoustic software, a \href{https://en.wikipedia.org/wiki/List_of_bioacoustics_software}{Wikipedia article} that lists bioacoustic software, and a compilation of \href{https://helpwiki.evergreen.edu/wiki/index.php/List_of_Bioacoustics_Software}{bioacoustic software} by the Evergreen State College.
Additionally, we explored several lists of annotation software on GitHub, including \href{https://github.com/heartexlabs/awesome-data-labeling}{heartexlabs/awesome-data-labeling}, \href{https://github.com/taivop/awesome-data-annotation}{taivop/awesome-data-annotation}, and \href{https://github.com/jsbroks/awesome-dataset-tools}{jsbroks/awesome-dataset-tools}.

To narrow down the list of potential alternative tools, we established specific criteria that each tool had to meet.
Firstly, the tool had to be available for installation or use through a web service.
We excluded any tool that require complex installation procedures or have outdated dependencies.
Secondly, the tool had to be user-friendly, which meant that it had to have a graphical user interface (GUI) and require no coding skills, as our goal was to identify tools that could be used by researchers with minimal technical expertise.
Thirdly, the tool had to be capable of visualizing audio files, either as a waveform or a spectrogram-like representation, as this is a fundamental aspect of bioacoustic annotation.
Fourthly, the tool had to provide a means for manual annotation and therefore only considered tools that were capable of generating annotations themselves.
Several bioacoustic tools provide automatic annotation capabilities, such as~\cite{szewczakSonoBat2010a}, but do not provide a means for manual annotation and therefore were excluded from our analysis.
Finally, we excluded any services that involved hiring external annotators, such as Amazon Mechanical Turk.
By applying these criteria, we were able to filter out tools that did not meet our requirements and narrow down our list of potential alternative tools to compare with \texttt{Whombat}.
In total, we evaluated 45 audio annotation tools.

The tools were then evaluated based on the following criteria:

\begin{itemize}
    \item[Open source] Whether the annotation tool is open source or not. Open-source tools allow users to access and modify the source code, which can be beneficial for researchers who need to customize the tool to fit their specific research needs.

    \item[Self-hosted] Whether the tool can be self-hosted, meaning it can be installed on a local server or personal computer and used without an internet connection. This is important for researchers who need to work with sensitive data that cannot be uploaded to a cloud-based platform, or are working under limited connectivity conditions.

    \item[Collaborative] Whether multiple users can use the tool at the same time. This is important for collaborative research projects where multiple annotators need to work on the same dataset simultaneously.

    \item[Large Datasets] This criterion evaluates the tool's ability to efficiently manage large datasets, enabling users to work with collections of recordings within a single workspace. Specifically, it assesses whether the tool enables users to browse quickly through multiple recordings without the need to manually load and unload each recording. While some tools, such as Raven \citep{yangCornellLabOrnithology2023} and Sonic Visualiser \citep{SonicVisualiser}, have the capability to load multiple recordings simultaneously, they may not be optimized for analysing large datasets.

    \item[Rich Metadata] Whether the tool can store and display rich metadata about the recordings. Many audio workstation tools, like Audacity \citep{audacity2017audacity}, do not display metadata about the recordings aside from the file name. Others, like Raven \citep{yangCornellLabOrnithology2023}, can display metadata about the recordings but do not allow the user to edit the metadata.

    \item[Search Capabilities] Whether the tool has search capabilities, allowing users to find specific annotations or recordings based on associated metadata. Search functionality is essential for efficient navigation and exploration of large datasets. It enables users to reference specific recordings or annotations quickly, improving the overall ease of use of the tool. Additionally, search capabilities enable users to filter recordings or annotations based on specific criteria, making it easier to identify unannotated files that should be included. 

    \item[Annotation Exploration] Whether the tool has annotation exploration capabilities. This means that the tool can display multiple annotations in a way that allows the user to visualize and compare several annotations simultaneously. In particular, we are interested in the ability to visualize annotations stemming from different recordings in the same workspace.

    \item[Flexible Spectrogram] Whether the tool has a flexible spectrogram generation system. This means that the tool can generate spectrograms with different parameters, such as the window size, the window type, the overlap, the colour scale, etc.

    \item[Flexible Annotation] Whether the tool has a flexible annotation system. This means that the tool can generate annotations of different types, such as point annotations, interval annotations, and bounding box annotations. Also, we require the ability to define custom tags, not restricted to species names or taxonomic terms.

    \item[Quality Assurance] Whether the tool includes integrated tools to help with quality control. These are any tools that help the user to check the quality of the annotations and flag potential errors.

    \item[Training tools] Whether the tool includes interactive components designed to assist in the training of novice annotators tailored to the current annotation objectives. Such components may include features that enable easy comparison of sounds to identify similarities and differences, or mechanisms to test an annotator's aural identification skills. Providing training tools can be especially useful for inexperienced annotators, allowing them to develop and refine their skills more quickly.

    \item[Prediction Evaluation] Whether the tool provides a mechanism for evaluating predictions against a set of ground truth annotations. Ground truth evaluation is essential for assessing the accuracy and reliability of automated annotation algorithms. By comparing the results of automated annotation against a known ground truth, it is possible to identify areas where improvements are needed.

    \item[Export Annotations] Whether the tool allows exporting the annotations into a sharable format with a clear schema. This is important for researchers who need to use the annotations in other software or for training machine learning models.

    \item[Integrated Detectors] This criterion evaluates whether the tool integrates automated detector capabilities. This means that the tool can use ML or otherwise to automatically generate annotations.
\end{itemize}

The evaluation of each tool was conducted by reading the documentation and user guides provided by the tool developers, or by using the tool itself when possible.
We acknowledge that this evaluation is subjective and that the results may be biased by the experience of the authors.

Out of the 45 tools evaluated, none met all the established criteria.
Notably, no tool included a component specifically designed to assist with annotator training, with the possible exception of tools that provided annotation instructions, such as~\cite{simpsonZooniverseObservingWorld2014}.
Only a small proportion of tools (less than 16\%) included features for quality control, annotation exploration, and prediction evaluation (Table~\ref{table:comparison_breakdown}).
These findings suggest that the majority of previously developed audio annotation tools did not prioritize the creation of annotated datasets suitable for machine learning development. To see the full list of tools evaluated consult the \texttt{annotation\_tool\_comparison.csv} file in the supplementary material.

\begin{table}[!ht]
\setcounter{table}{0}
\renewcommand{\thetable}{SI.\arabic{table}}
\begin{center}
\begin{tabular}{lcc}
                        & \multicolumn{1}{l}{Total} & \multicolumn{1}{l}{Percentage} \\
Open Source             & 28                        & 62.2\%                        \\
Self Hosted             & 41                        & 91.1\%                        \\
Collaborative Use       & 13                        & 28.9\%                        \\
Handling Large Datasets & 23                        & 51.1\%                        \\
Rich Metadata Display   & 11                        & 24.4\%                        \\
Search Capabilities     & 9                         & 20.0\%                        \\
Annotation Exploration  & 7                         & 15.6\%                        \\
Flexible Spectrogram    & 26                        & 57.8\%                        \\
Flexible Annotation     & 4                         & 8.9\%                         \\
Quality Control         & 5                         & 11.1\%                        \\
Annotator Training      & 0                         & 0.0\%                         \\
Prediction Evaluation   & 1                         & 2.2\%                         \\
Integrated Detectors    & 14                        & 31.1\%                       
\end{tabular}
\caption{\label{table:comparison_breakdown} Comparison of the tools evaluated based on the established criteria.
    The table presents the number of tools that satisfied each criterion evaluation and the percentage of tools that satisfied each criterion.}
\end{center}
\end{table}

\newpage
\section{Software Design}
\label{appendix:software_design}

\texttt{Whombat} was designed with usability, scalability, and extensibility as priorities.
Here we outline the key design decisions we made and the rationale behind them.

We believe that open source software fosters collaboration, innovation, and transparency.
Therefore, we decided to release our audio annotation tool as an open source project on a public repository.
This allows other researchers, developers, and users to access, use, modify, and contribute to our codebase.
We also provide documentation, examples, and tutorials to facilitate the adoption of our tool by the community.

We opted for a server-client configuration for our audio annotation tool as it affords the flexibility to host both backend and frontend on either separate machines or one machine, depending on resource availability and utilization.
This approach also facilitates exploiting web-based interface advantages such as portability and accessibility.
In particular, we created the backend with a RESTful API that manages communications between client-server requests/responses while executing audio processing pipelines and saving outputs in a database.
On the other hand, the frontend is responsible for displaying meaningful data to users as well as handling their interactions with it.

We chose \texttt{Python} \citep{van1995python} as the main language for the backend of our audio annotation tool because of its rich ecosystem of packages for scientific computing, data analysis, and web development.
Our preference towards utilizing \texttt{Python} also enables seamless integration with multiple machine learning tools and pipelines available for \texttt{Python}.
Furthermore, we observed that using \texttt{Python} facilitates code sharing and collaboratation due to its learnability and readability.
\texttt{Python} has a large and active community of developers, researchers, and enthusiasts, which can provide support and feedback.

To implement our RESTful API, we employed the \texttt{FastAPI} \citep{ramirezFastAPI} framework for its lightweight and flexible characteristics.
For audio processing tasks, we utilized the \texttt{scipy} \citep{virtanenSciPyFundamentalAlgorithms2020} and \texttt{numpy} \citep{harrisArrayProgrammingNumPy2020} packages, which provide a wide array of functions for scientific computing and data analysis.
All data produced and used by our audio annotation tool is saved in a relational database.
\texttt{SQLite} was our default choice of database management system due to its lightness and efficiency in managing small to medium-sized datasets.
Nonetheless, we acknowledge that certain users may require other database systems, such as \texttt{MySQL} or \texttt{PostgreSQL}, contingent on their specific needs and constraints.
Thus, we offer a configuration option to enable switching to a different database backend according to the user's preference.
Communication between the database and the backend was facilitated by the \texttt{SQLAlchemy} \citep{mikebayerSQLAlchemy2023} package, providing a high-level interface for managing database systems.
Moreover, we provide a \texttt{Python} API that enables direct interaction with the stored data, allowing users to create customized analysis pipelines or integrate data into machine learning pipelines.
This feature provides flexibility and extensibility beyond the tool's default functionality.

In selecting a language for the user-facing components of our audio annotation tool, we opted for \texttt{TypeScript} \citep{biermanUnderstandingTypeScript2014}, a superset of \texttt{JavaScript} that includes optional static typing to enhance code quality.
For constructing the interface itself, we turned to \texttt{React} \citep{walkeReact2023}, a widely used and effective library that employs a declarative and component-based approach to building interfaces.
This approach affords us greater consistency in design and allows us to reuse UI elements.

We wrote the audio annotation tool with the aim of making it easy to understand and extend.
To that end, we added comprehensive documentation in all the main modules and functions, including detailed explanations of the inputs, outputs, and behavior of each component.
We also provided examples of how to use the tool in practice, as well as clear instructions for setting up and configuring the tool.
Additionally, we implemented unit and integration tests in the most critical parts of the software, to ensure correct behavior and facilitate future development.
These tests cover a wide range of scenarios and edge cases, and are automatically run whenever changes are made to the codebase.
By providing clear documentation and robust testing, we hope to make it easier for users to understand and extend our tool, as well as contribute to the broader bioacoustics community.

\end{document}